\documentclass[american,a4paper]{article}
\usepackage[%
  checkReferences=false]{packages}

\graphicspath{
  {./plots},{plots}
}
\usepackage{macros}

\usepackage{authblk}
\title{The spectrum of light isovector mesons with $C=+1$ from the COMPASS experiment}
\author{
Stephan \textsc{Paul} 
	\\ for the COMPASS collaboration
	}
\affil{Physik Department, Technical University Munich, James Franck Str., 85748 Garching, Germany\\ {\rm email:} \href{mailto:stephan.paul@tum.de}{stephan.paul@tum.de}}
\date{September 30, 2016}

\begin{document}
\maketitle

\begin{abstract}
Based on the largest event sample of diffractively produced \threePi, obtained by a pion beam of \SI{190}{\GeVc} momentum, the COMPASS collaboration has performed the so far most advanced partial-wave analysis on multi-body final states, using the isobar model. The large number of 88waves included in the analysis reduces truncation effects. We have used fourteen waves, to extract resonance parameters for eleven light-meson candidates, most of them observed previously. The coherence of the analysis and the large variety of systematic studies has allowed us to determine mass and width of most $a_{J}$ and $\pi_{J}$ states with a total of six different values of \JPC below a mass of \SI{2.1}{\GeVcc}, with high confidence. We exploit that the production rates of resonant and non-resonant contributions in these fourteen waves vary differently with the four-momentum transfer squared in the reaction. In addition, we have performed the first isobar-freed analysis in diffraction, from which we have determined the shape of the \pipiSW isobar for different \JPC of the $3\pi$ system.
\end{abstract}

%\begin{keywords} 

\section{Introduction}
\label{sec:introduction}
The excitation spectrum of light-quark bound states has gained much
interest in the last years.  Recently, the
simulation of QCD on the lattice has caught new momentum because it
now also addresses the dynamics of meson decays, which will lead to
more realistic predictions for masses and widths of excited hadrons.
Thus, a precise knowledge of the spectrum of light hadrons has become
important. Excited light-quark hadrons occur in the
decay of heavy-quark mesons and are currently studied extensively in
high-flux scattering experiments at CERN~\cite{Abbon:2014aex} and
JLAB~\cite{Battaglieri:2010zza,Ghoul:2015ifw}.
At present, results from different experiments, summarized by the Particle Data Group (PDG)~\cite{Agashe:2014kda}, vary considerably or even are
inconsistent. Similarly, the interpretation of many states is controversial, as is the case e.g. for the new
axial-vector state \PaOne[1420] observed by COMPASS~\cite{Adolph:2015pws}, which appears with 
the same quantum numbers as the elusive \PaOne \cite{Aceti:2016yeb,Wang:2015cis,Chen:2015fwa,Ketzer:2015tqa,Wang:2014bua}.
Mesons are characterized by their quantum numbers, isospin $I$ and \JPC, with $J$ being the total spin, $P$ the parity and $C$ the charge conjugation quantum number\footnote{Although the $C$
  parity is not defined for a charged system, it is customary to quote
  the \JPC quantum numbers of the corresponding neutral partner state
  in the isospin multiplet. The $C$ parity can be generalized to the
  $G$ parity $G \equiv C\, e^{i \pi I_y}$, a multiplicative quantum
  number, which is defined for the non-strange states of a meson multiplet.}. 
 Extensive discussions of the light-meson
sector are found in~\refsCite{klempt:2007cp,Brambilla:2014jmp}.
Here, we shall restrict ourselves to isovector states
with assigned
positive $C$ parity and masses below about \SI{2.1}{\GeVcc}, decaying into \threePi.
%three charged pions. 

%The COMPASS collaboration has already studied properties of isovector
% $3\pi$ resonances~\cite{adolph:2014mup,alekseev:2009aa} in the mass
% range between \SIlist{1.1;2.1}{\GeVcc} using a lead target.  In a
% previous publication~\cite{Adolph:2015tqa}, we have presented the
% study of isovector mesons decaying into three charged pions using a
% hydrogen target with the emphasis on \one production kinematics, \two
% {\color{blue}distinction} of non-resonant processes
% \todo{separation: did we show this?},  \three search for new and excited mesons {\color{blue}as the \PaOne[1420]}
%, and \four on properties of the \pipiSW amplitude.  This paper is the third\todo{check} in a planned series of publications to present
%precision {\color{blue}spectroscopy} revisiting all quantum numbers accessible in reaction~\eqref{eq:reaction} up to total spin $J = 6$.
% The analysis is limited to states belonging to the family of $\pi_J$ and $a_J$.
%\todo{rework this paragraph; most sentences probably have to be moved up}

This work is based on the currently world's largest data of $50\cdot 10^{6}$ events set on
diffractively produced mesons decaying into \threePi, which has
previously been discussed in~\refCite{Adolph:2015tqa}.

\section{Diffraction and Partial-Wave Analysis}
\label{sec:PWA}
\subsection{Diffraction}
\label{sec:diffraction}
\begin{wrapfigure}{r}{0.48\textwidth} 
\vspace{-20pt}
  \begin{center}
%\begin{figure}[htbp]
%  \centering
  \includegraphics[scale=0.95]{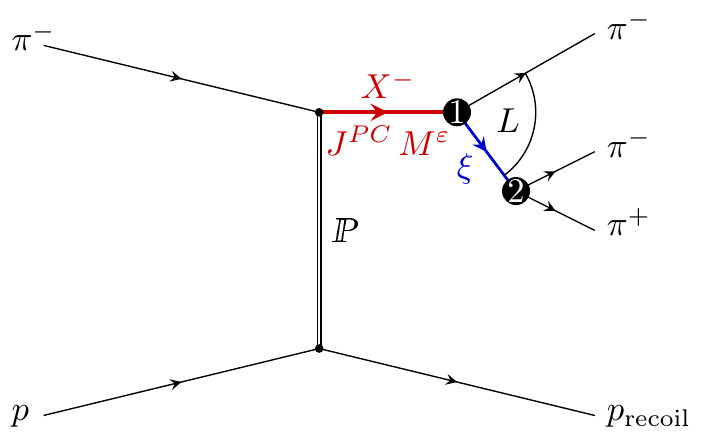}
  \caption{Diffractive production of $X^-$ with quantum numbers \JPC and spin projection $M$, and naturality $\varepsilon$ of the exchange particle. The decay is described in the isobar model and is assumed to proceed via an intermediate \twoPi state $\xi$, the so-called isobar. The relative orbital angular momentum of the isobar and the spectator pion is denoted by $L$.}
  \label{fig:3pi_reaction_isobar}
%\end{figure}
 \end{center}
  \vspace{-30pt}
  \vspace{1pt}
\end{wrapfigure}

Information on light hadrons can be obtained from heavy-meson decays
and \pbarp annihilations where they appear in subsystems of the
measured final state and can be identified as interfering amplitudes
in the Dalitz-plot.  In diffractive production with incoming
negatively charged pions, the beam particle is excited by the
exchange of a Pomeron with the target nucleon or nucleus to form
short-lived intermediate states $X^-$ (see \cref{fig:3pi_reaction_isobar}). The detection of the recoil proton suppresses events with inelastically scattered target particles. Still, more complex exchange processes may occur, which are not described by \cref{fig:3pi_reaction_isobar} but produce the same final state without passing through a resonance. These
processes form a coherent background for resonance production and have to be singled out by other means.
The decay of the states $X^-$ proceeds independently of its production, which therefore factorize.
The various $X^-$ contributing coherently can be separated using
partial-wave analysis (PWA).  
\subsection{Partial-wave analysis}
\label{sec:partial-wave-analysis}
For this, we employed  the
isobar model. It describes the transition of $X^-$ into the final
state (here \threePi) as a sequence of two-body decays via additional
intermediate states $\xi$, called isobars. The decays are represented by the vertices labeled~1 and~2 in \cref{fig:3pi_reaction_isobar}. 
%and are mathematically described by their corresponding decay amplitudes.  
Both $X^-$ and $\xi$ are
characterized by their quantum numbers \JPC. Together with the spin projection $M$ of $X^-$ and the reflectivity quantum number $\epsilon$ (see~\refCite{Adolph:2015tqa}) it defines a wave, which represents a characteristic pattern in the five-dimensional phase space of the \threePi final state and also includes non-resonant processes. The full data set contains a
mixture of coherent and incoherent contributions of various waves.
These contributions (complex-valued production amplitudes) are extracted by
an extended maximum likelihood fit. 
%
%\vspace{-.5cm}
%
\begin{figure}[htbp]
  \centering
  \includegraphics[scale=0.55]{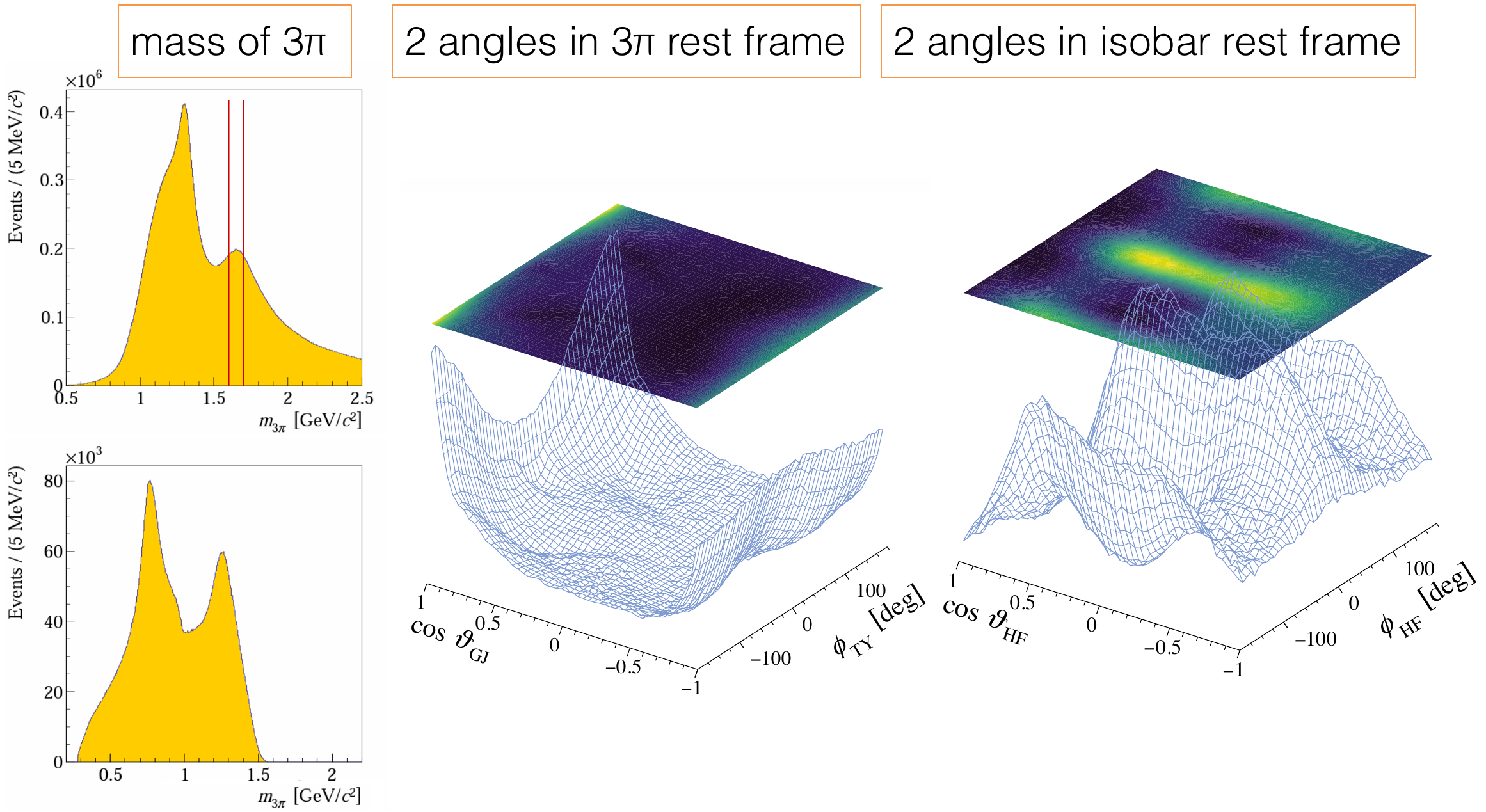}
  \caption{Kinematic distributions: spectrum of the invariant mass \mThreePi of the final state (upper left); mass spectrum of the $\pi^{+}\pi^{-}$ subsystem (lower left) and the 2-body angular correlations in the Gottfried-Jackson frame (centre) and helicity frame (right), the latter three for $\mThreePi\in\SIrange{1.6}{1.7}{\MeVcc}$ .}
  \label{fig:raw_data}
\end{figure}
%\vspace{-.5cm}
%
Such a fit is performed independently for individual bins of mass of
the $3\pi$ final state \mThreePi, subdivided into bins in the reduced four-momentum transfer \tpr. The fits result in a spin-density matrix for each
bin of mass and \tpr.  A detailed description can be found in
\refCite{Adolph:2015tqa}.  Here, \tpr is defined by
%
%\vspace{-0.2cm}
\begin{equation}
 \tpr \equiv \tabs - \tmin \geq 0, ~\text{where}~\tmin \approx \rBrk{\frac{\mThreePi^2 - m_\pi^2}{2 \Abs{\vec{p}_\text{beam}}}}^2~~,
% \label{eq:tmin}
\end{equation}
\vspace{-0.1cm}
\noindent where $t$ is the four-momentum transfer squared, $m_{\pi}$ the pion mass and $p_\text{beam}$ the beam momentum.
%\todo{mention \tpr bins}.  
%
%
%\vspace{-.5cm}
\begin{figure}[h]
  \centering
  \includegraphics[scale=0.6]{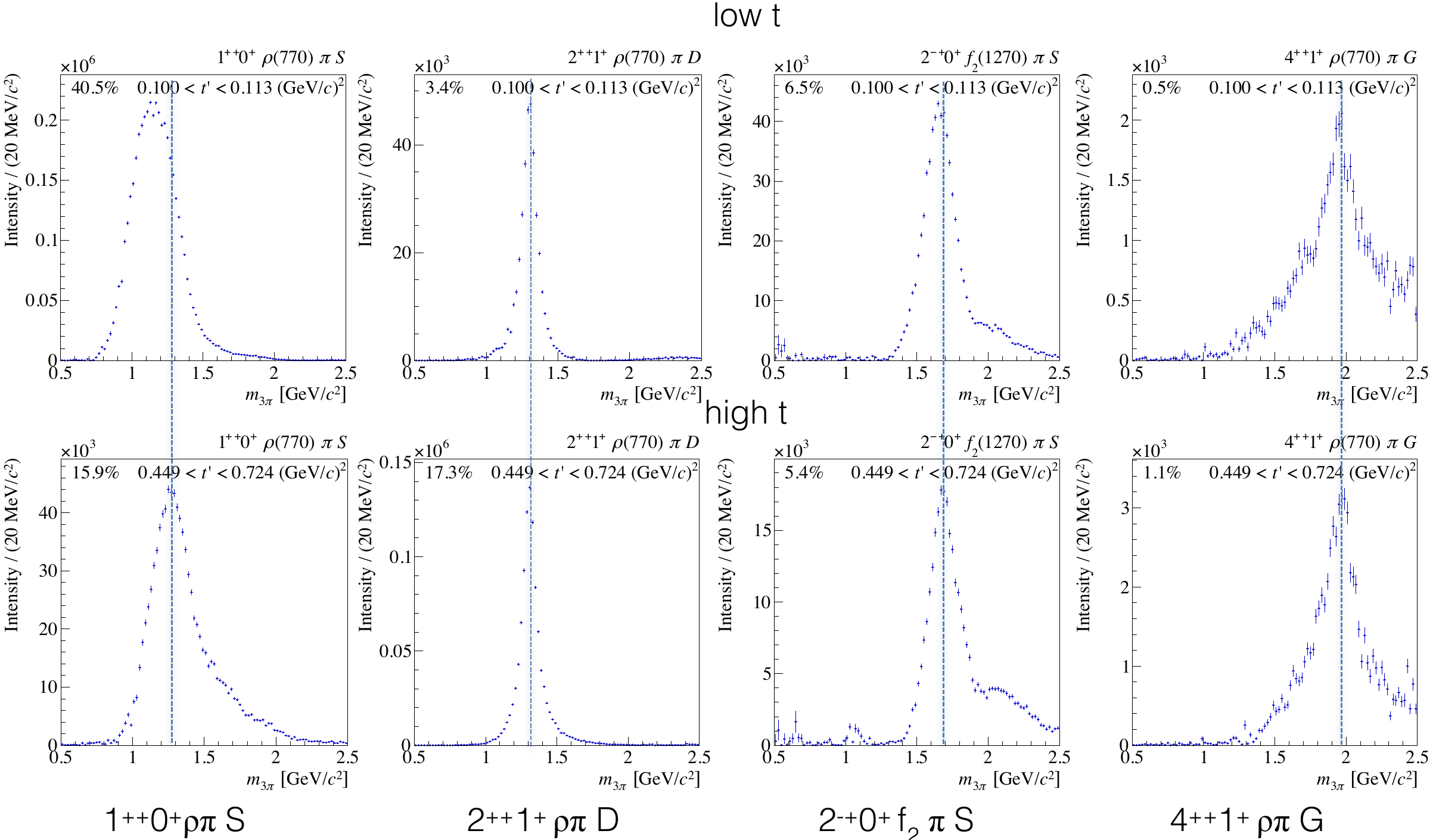}
  \caption{Intensity spectra for individual waves resulting from partial-wave fits. Each data point corresponds to the result of an independent fit. We show the spectra for two bins of \tpr (see text for details). Vertical lines are placed to guide the eye and indicate the intensity maximum for each wave at high values of \tpr. The fit fractions for the particular bin in \tpr are indicated for each wave.}
  \label{fig:mass-independent-fit}
%  \vspace{-.95cm}  
\end{figure}
\par
\Cref{fig:raw_data} shows some characteristic distributions for the \threePi final state: the spectrum of the final state mass \mThreePi, the mass distribution for the \twoPi subsystem for $\mThreePi\approx\SI{1.6}{\GeVcc}$ and the corresponding angular distributions in the Gottfried-Jackson frame of $X^{-}$ and the helicity frame of the isobar. The obvious dominance of isobars appearing in the \twoPi subsystem motivates the use of the isobar model in our analysis. The key issue of this first step the PWA analysis is the choice of waves forming a set. COMPASS has used the largest set ever, consisting of a coherent sum of 80 waves with positive reflectivity (natural parity exchange in the production process), a coherent sum of seven waves with negative reflectivity and one single incoherent wave representing pure phase space. About 97\% of the total intensity can be described by waves with distinct isobaric character.
\par
\noindent 
The result of this analysis is shown exemplarily in \cref{fig:mass-independent-fit} for four prominent waves. The wave nomenclature follows $\JPC M^{\epsilon}[\textrm{isobar}]\pi L$. Each data point corresponds to the result of a single fit. We show the spectra for two bins of \tpr, the lowest bin of \tpr, $0.100 < \tpr < 0.113~(\rm{GeV}/\rm{c})^{2}$, and the second highest one, $0.449 < \tpr < 0.724(\rm{GeV}/\rm{c})^{2}$.
The spectra demonstrate the interplay of resonant and non-resonant components, whose production amplitudes exhibit different \tpr dependences. In particular, for the waves with $\JPC=\onePP$ and $\JPC=\twoMP$ the apparent peak position moves with \tpr, while the spectra for $\JPC=\twoPP$ and $\JPC=\fourPP$ show very little dependence on \tpr. This effect is independent of the charge combinations of the $3\pi$ final state, as is indicated for the case of $\JPC=\onePP$. Performing the partial-wave decomposition independently in narrow bins of \tpr allows us to better disentangle resonant and non-resonant amplitudes in the second analysis step (see \cref{sec:resonance-fits}).

%-----------------------------------------
\section{Resonance-model fits}
\label{sec:resonance-fits}
\begin{figure}[htbp]
  \centering
  \includegraphics[scale=0.73]{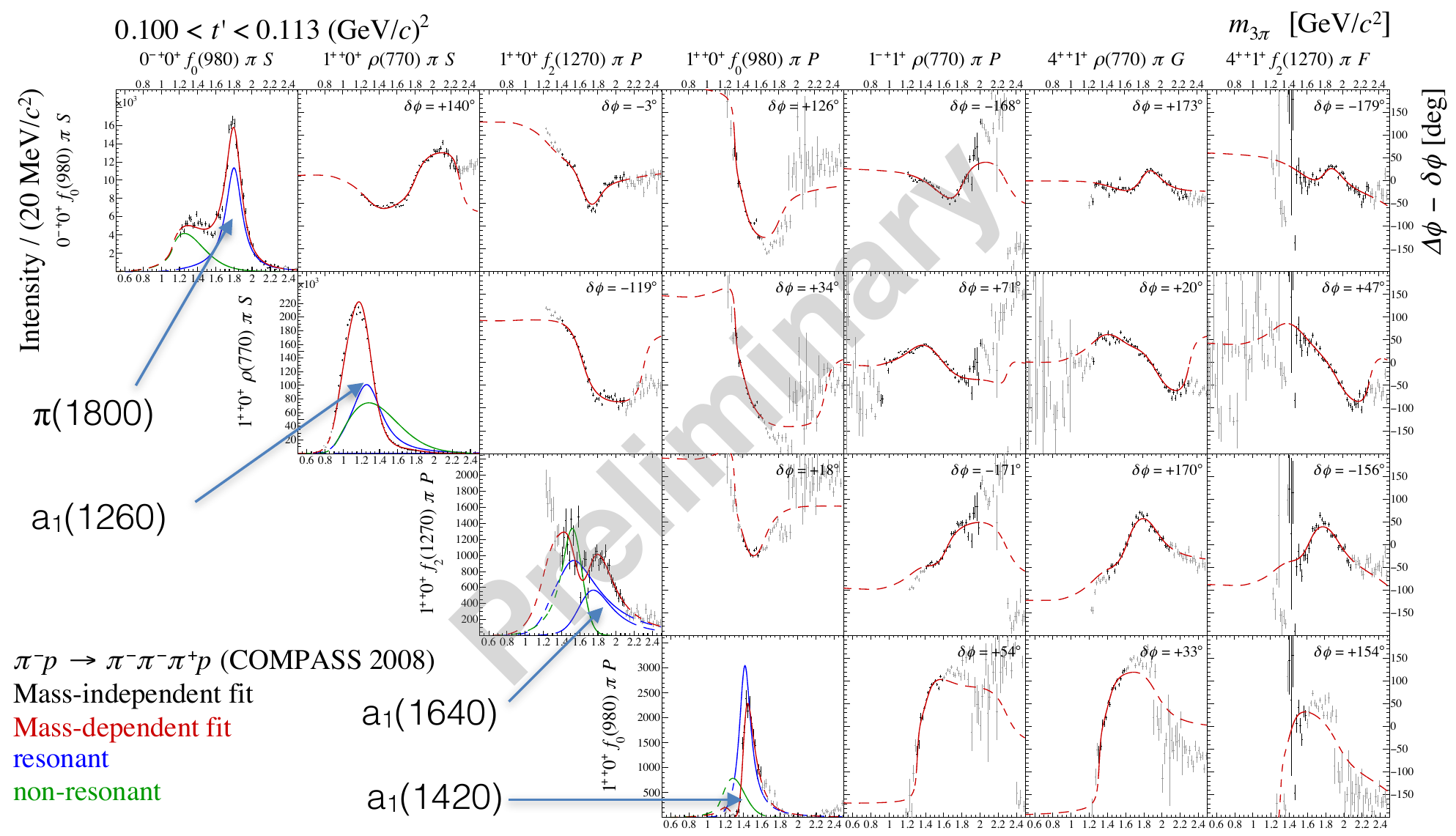}
  \caption{Submatrix of the full spin-density matrix obtained from the first step of our PWA. We show intensities for the $\JPC=\zeroMP \textrm{and}~\onePP$ waves used in our 14-wave fit, and the relative phases with respect to seven waves. The green curves represent the non-resonant contributions, the blue curves the resonances. The red solid lines represent the coherent sum of all contributions, the dotted part of the red line is the extrapolation into mass regions outside the fit range. The blue arrows point to resonances, assigned to the closest matching PDG entry~\cite{Agashe:2014kda}.}
  \label{fig:spin-density-matrix-a1}
\end{figure}

The steps described in \cref{sec:partial-wave-analysis} are the
prerequisite to search for resonances produced in the reaction
process, which can only be identified if we combine the information from
all spin-density matrices over a wide range of final-state masses.
For this step, different methods were used in the past. This includes
$K$-matrix formalisms, where resonances are described by poles and
the description of resonances by Breit-Wigner (BW) amplitudes. Here, we only treat one final state, namely \threePi, and thus the full $K$ matrix cannot be obtained from the data alone. In addition, the description of the non-resonant processes lacks a stringent recipe, and thus the fit model for this work consists of a sum of a set of BW amplitudes to describe resonances and a set of exponentials in the two-body break-up momentum for the background, separately.  This technique was employed in previous analyses including our own work
\cite{Adolph:2015pws,alekseev:2009aa}. The major extension of the number of waves allows to identify almost the complete isovector family with $C=+1$ at once and to determine systematic uncertainties in a consistent way. Theoretical and technical work necessary to implement $K$ matrices is under development.
%\vspace{-0.5cm}
\noindent The fit of the full spin density matrix composed of positive reflectivity waves only is technically very difficult owing to the large number of free parameters and requires a good modeling of the non-resonant contributions for all waves. Remaining artifacts from either the truncation of the partial-wave series or inherent to the isobar model render this task impractical and thus we have restricted our fits to only 14 waves listed in \cref{tab:rmf:waveset}.
%containing $\JPC=\zeroMP, \onePP, \oneMP, \twoPP, \twoMP \textrm{and} ~\fourPP$. 
We present and discuss the results in the following sections. 
%\vspace{-0.5cm}
%
%
\begin{wraptable}{r}{0.55\textwidth} 
%\vspace{-0.3cm}
%
%\begin{table}
	\centering	
	\vspace{-5pt}

	\caption{Waves used in the resonance-model fit}
	  	\vspace{6pt}	
		
	\label{tab:rmf:waveset}
	\begin{tabular}{|l|l|}
	\hline
 		\wave{0}{-+}{0}{+}{\PfZero}{S}   &\wave{1}{-+}{1}{+}{\Prho}{P} \\
		\wave{1}{++}{0}{+}{\Prho}{S}      & \wave{1}{++}{0}{+}{\PfTwo}{P}  \\                                                                                   
		\wave{1}{++}{0}{+}{\PfZero}{P} & \wave{2}{++}{1}{+}{\PfTwo}{P}  \\                                                                                
       		 \wave{2}{++}{2}{+}{\Prho}{D}   & \wave{2}{++}{1}{+}{\Prho}{D} \\
		 \wave{2}{-+}{0}{+}{\PfTwo}{S}  & \wave{2}{-+}{0}{+}{\PfTwo}{D} \\
		 \wave{2}{-+}{0}{+}{\Prho}{F}    & \wave{2}{-+}{1}{+}{\PfTwo}{S}  \\
		 \wave{4}{++}{1}{+}{\Prho}{G }  & \wave{4}{++}{1}{+}{\PfTwo}{F}  \\ \hline
	\end{tabular}
%\end{table}
\end{wraptable} 
%\vspace{-0.5cm}
% \vspace{-30pt}
  \vspace{-2pt}
\noindent \Cref{fig:spin-density-matrix-a1} shows part of the spin-density submatrix used in the fit for one interval of \tpr. It relates the waves with $\JPC=\zeroMP \textrm{and}~\onePP$ to six of the other waves used. The fit model uses one relativistic BW each for \zeroMP, \oneMP, \fourPP and \wave{1}{++}{0}{+}{\PfZero}{P}, two BW for \wave{1}{++}{0}{+}{\Prho}{S} and \wave{1}{++}{0}{+}{\PfTwo}{D} as well as $\JPC=\twoPP$ and three BW for the three waves with $\JPC=\twoMP$. The non-resonant contributions are parametrized using exponentials of the same shape, with the exception of five waves for which the shape is \tpr dependent and modified by a power law. In order to avoid local maxima in the parameter space of the $\chi^{2}$ function, we make 1000 fit attempts with randomly sampled starting values for the fit parameters and use different schemes for releasing and fixing of the parameters during the fit. Details of the fit procedure will be published soon in a dedicated paper.
\subsection{Fit results}
\label{sec:fit_results}
\begin{figure}[htbp]
  \centering
  \includegraphics[scale=0.37]{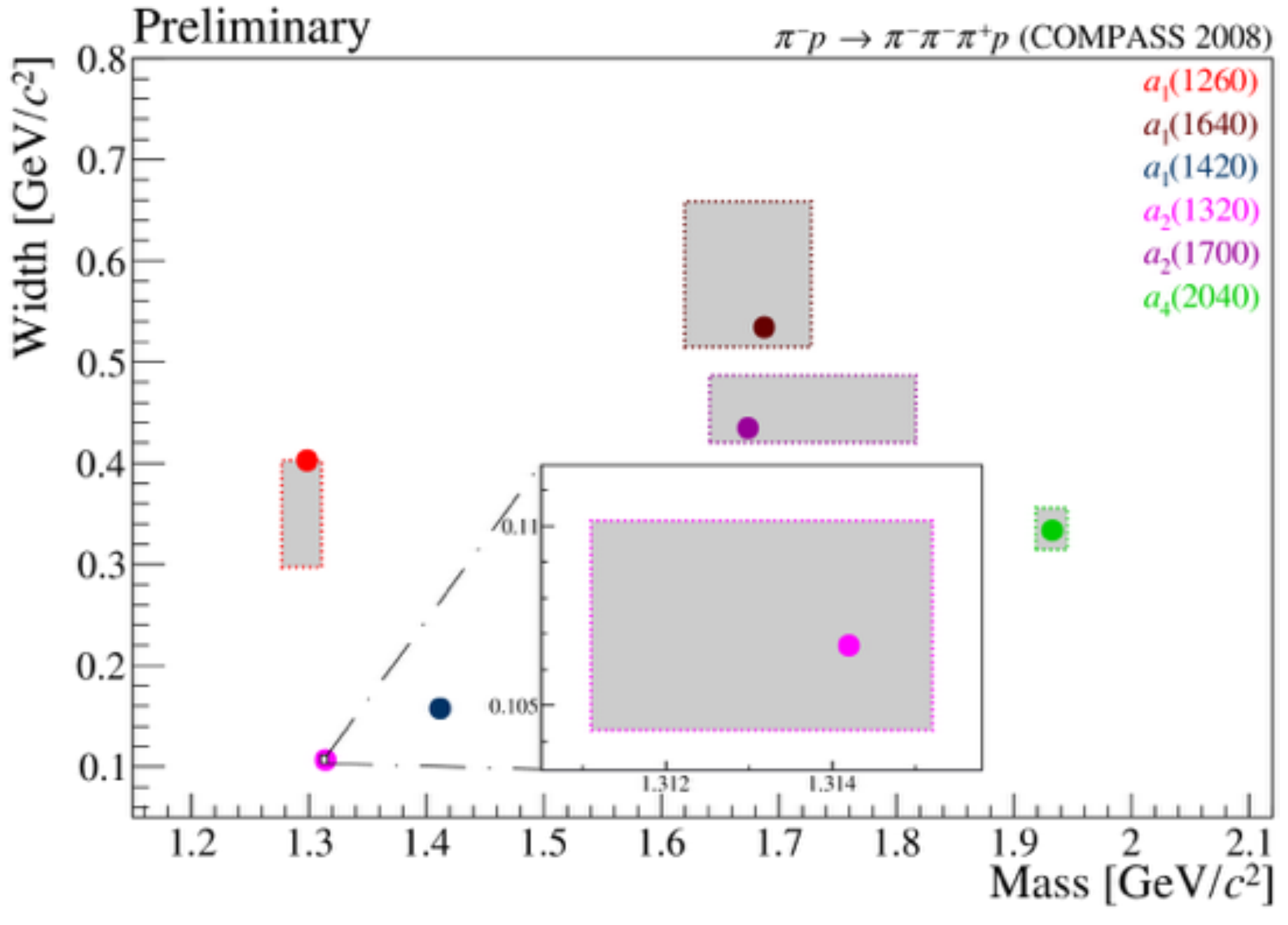}
  \includegraphics[scale=0.37]{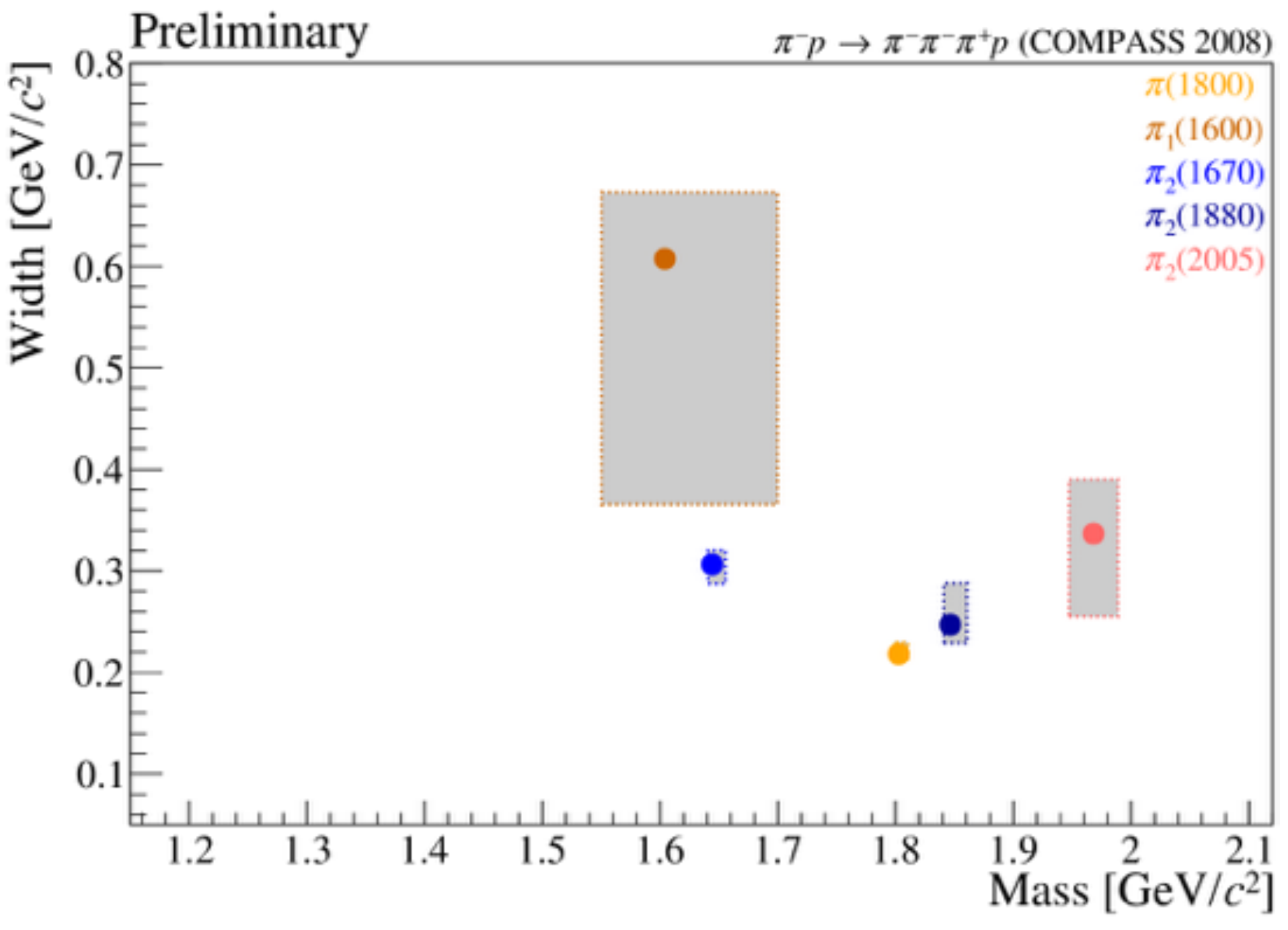}
  \caption{Results of the systematic studies for the resonance parameters of $a_{J}$ states (left) and $\pi_{J}$ states (right). The horizontal axes show the BW mass, the vertical axes the width of the BW. The size of the box indicates the systematic variations from a total of up to 20 systematic studies.}
  \label{fig:fit-summary}
\end{figure}

The results for all resonance are visualized in \cref{fig:fit-summary}. For $\JPC=\onePP$, we observe three resonances. They appear with different relative strength in the various waves and thus allow a parametrization in terms of BW resonances. The strong \wave{1}{++}{0}{+}{\Prho}{S} wave contains a dominant structure around \SI{1.2}{\GeVcc}, which we decompose into a large non-resonant contribution and the \PaOne. This decomposition is supported by the different \tpr dependence of the components. However, there is no reliable interferometer at low masses, which would allow to observe the \SI{180}{\degree} phase variation of \PaOne. Thus, resonance parameters show large uncertainties, but we can rule out large values for the width of  \PaOne as reported by previous experiments. \PaOne[1640] is hidden in this wave, but it becomes visible in the phase variation with respect to other waves and is distinctly observed in the spectral distribution of \wave{1}{++}{0}{+}{\PfTwo}{P}. As this resonance appears in a mass region populated by many resonances with other \JPC, the extracted BW parameters are sensitive to choices in the analysis model. \PaOne[1420] was previously unknown and could be unraveled in the \wave{1}{++}{0}{+}{\PfZero}{P} wave, despite its small production strength. The resonance parameters are very robust towards any systematic variations of analysis parameters.
%
%\vspace{-0.5cm}
\noindent For $\JPC=\twoPP$, we observe two states, \PaTwo and \PaTwo[1700]. The latter exhibits strong interference effects with the non-resonant contribution in \wave{2}{++}{1}{+}{\Prho}{D}, which change from destructive to constructive as function of \tpr. The mass spectrum of \wave{2}{++}{1}{+}{\PfTwo}{P} shows a clear distinction of the two $a_{2}$ states.  While the determination of the resonant parameters for \PaTwo is unambiguous, we observe large systematic uncertainties for \PaTwo[1700], owing to the large width of this state.

\noindent We observe three states with $\JPC=\twoMP$, which are relatively close in mass and have widths around \SIrange{250}{350}{\MeVcc}. Similarly to the $a_{1}$ sector, these resonances appear in the various waves with different strength.  The \wave{2}{-+}{0}{+}{\PfTwo}{S} and \wave{2}{-+}{1}{+}{\PfTwo}{S} waves are dominated by \PpiTwo, which, however, is very weak in \wave{2}{-+}{0}{+}{\PfTwo}{D}. The latter wave in turn is dominated by \PpiTwo[1880] with a smaller contribution of \PpiTwo[2005]. The wave \wave{2}{-+}{0}{+}{\Prho}{F} shows a clear signal for \PpiTwo[2005], in particular at large values of \tpr. Also \PpiTwo appears here, while \PpiTwo[1880] plays a minor role. We conclude from our analysis that \PpiTwo[1880] is an independent state and thus disfavor the suggestions of it being a reflection of \PpiTwo interfering with non-resonant contributions~\cite{dudek:2006ud}. Our systematic studies reveal correlations in the determination of BW parameters for the two excited $\pi_{2}$. 

\noindent The spin-exotic state \PpiOne is hidden beneath a large non-resonant contribution at small values of \tpr. However, similarly to \PaOne, it can be singled out at our largest values of \tpr. While the determination of the mass of \PpiOne is rather robust, the value for the width extracted from the various fits reveals large uncertainties, mostly connected to the sector of $\JPC=\onePP$. The exotic \PpiOne appears wider than previously observed. For $\JPC=\zeroMP \textrm{and}~\fourPP$, we observe the well known states \Ppi[1800] and \PaFour and determine the resonance parameters with very small systematic uncertainties. The values for mass and width are consistent with previous observations, however, \PaFour is slightly lighter than determined by ~\refCite{Agashe:2014kda}, using several previous measurements.

\begin{figure}[htbp]
  \centering
  \includegraphics[scale=0.53]{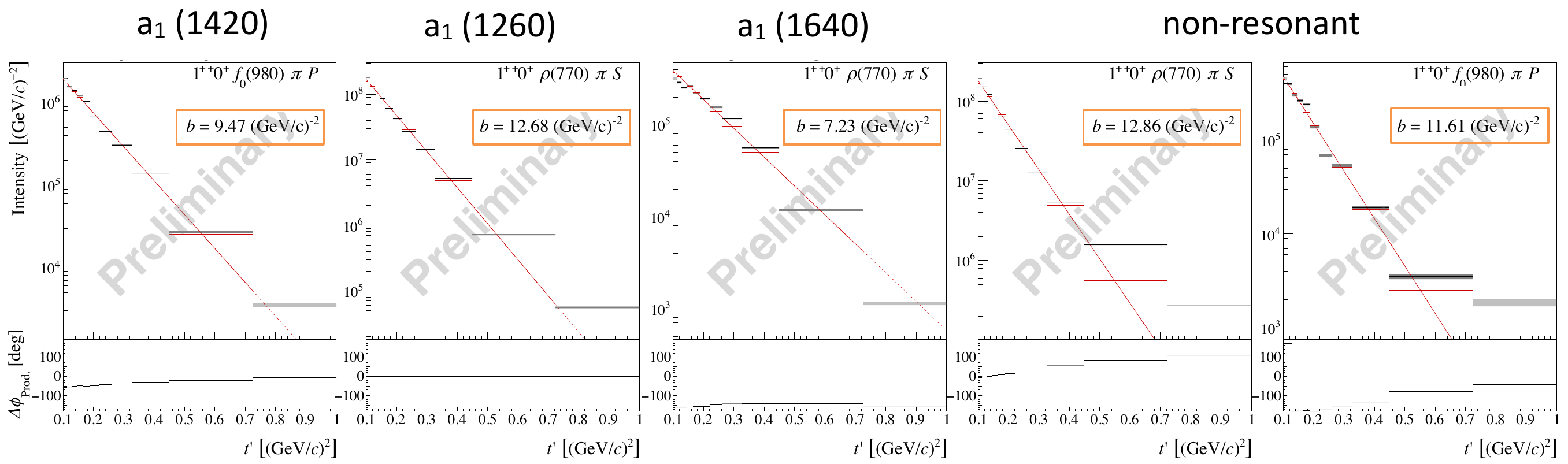}
  \caption{\tpr dependence of the integrated yield for the different fit components for $\JPC=\onePP$ (upper row). \tpr dependence of the production phase \wrt to \PaOne for the different components (lower row).}
\label{fig:tpr_dependence}
\vspace{-0.2cm}

\end{figure}

%\vspace{-0.3cm}
\noindent The separation of various components within all waves has strongly profited from their different \tpr-dependent production rates. We have integrated their contributions over \mThreePi and the resulting \tpr dependence has been described by a single exponential multiplied by $\tpr^{|M|}$, with $M$ being the spin projection quantum number. The result is shown exemplarily for $\JPC=\onePP$ in \cref{fig:tpr_dependence}. The slope parameter $b$ of the exponential shows some regular patterns, although with some exceptions. In general, the slope of the \tpr dependence flattens with increasing mass of a resonance within a particular \JPC. Most values for $b$ are in the range of \SIrange{6}{9}{\perGeVcsq} for resonances and non-resonant contributions often drop much faster than the resonant ones. Some exceptions are the slope parameter for \PaOne of $b=\SI{12.7}{\perGeVcsq}$ or the very shallow slopes for the non-resonant contributions for $\JPC=\twoMP$, where these non-resonant contributions are small and we fit with three BW functions. Similar is true for $\JPC=\twoPP$, where again the non-resonant contributions are small and may not be fully separated from the resonant ones.

\section{Freed-Isobar Analysis}
\label{sec:deisobarred}
The conventional isobar model implies that final-state interaction does not alter the shape of the isobars, which are implemented with a fixed shape taken
from~\refCite{Agashe:2014kda} or, for the \pipiSW, from $\pi\pi$ scattering data. In order to test this hypothesis, 
%(and on the way to replace the classical isobar Ansatz), 
we have replaced the fixed parametrization of \pipiSW by a series of step-like functions across the \twoPi mass spectrum, individually for each bin of \mThreePi. Owing to the large increase in the number of free parameters, we have restricted this exercise to only three values of \JPC for \threePi with \pipiSW isobar, namely \zeroMP, \onePP and \twoMP. 
%\vspace{-0.5cm}

\begin{figure}[h]
  \centering
  \includegraphics[scale=0.65]{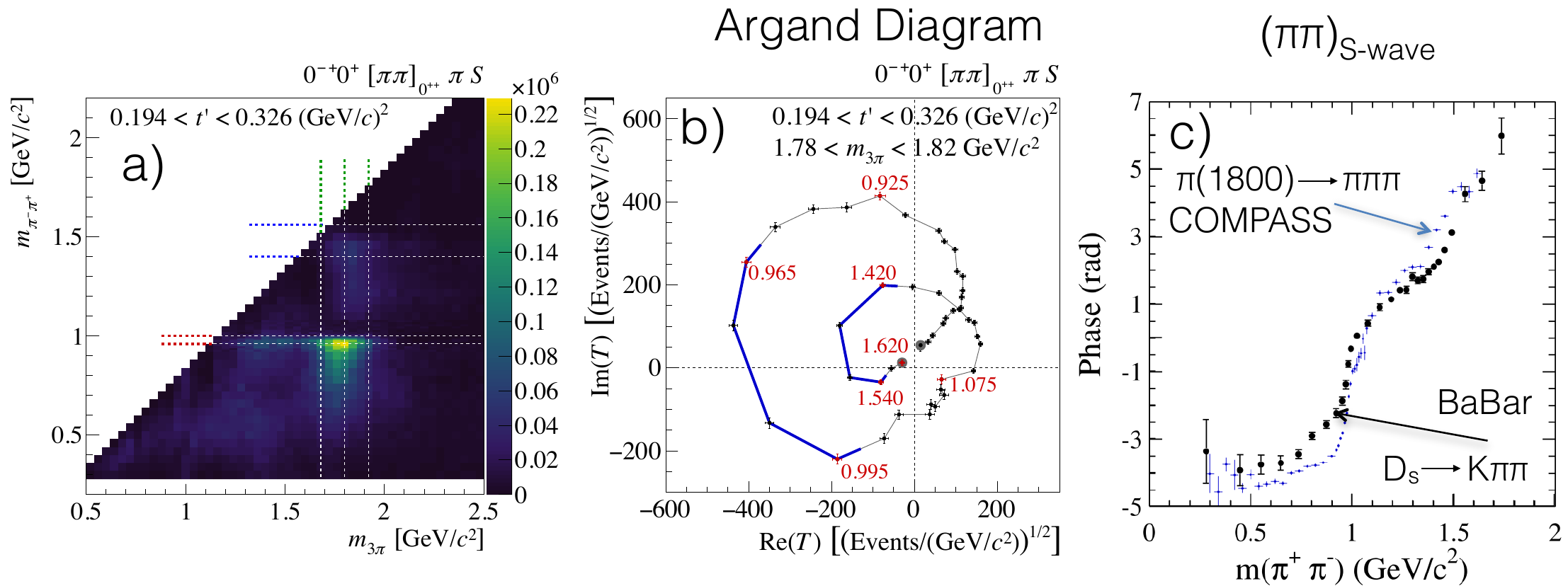}
  \caption{Results from a freed-isobar PWA. a) \mTwoPi in \zeroPP vs. \mThreePi in \zeroMP for one bin in \tpr; b) Argand diagram for the corresponding $\pi^{-}\pi^{+}$ amplitude selecting \mThreePi  around the mass of \PpiZero[1800]. Red labels indicate \mTwoPi along the trajectory; c) comparison of the \pipiSW in three body decays from $D_{s}$ (Babar~\cite{Aubert:2008ao}) and \PpiZero[1800] from this analysis.}
  \label{fig:deisobarred}
\end{figure}
%\vspace{-0.5cm}

%
The result is a correlation of the spectral distribution of the $\pi^{-}\pi^{+}$ system with $\JPC=\zeroPP$ and the corresponding \mThreePi spectrum with definite \JPC. \Cref{fig:deisobarred}a shows the result for the \zeroMP system. As we extract the complex amplitude for the $\pi^{-}\pi^{+}$, we present in \Cref{fig:deisobarred}b the mass dependence of real and imaginary part in an Argand diagram for a \mThreePi interval around \PpiZero[1800]. With increasing \mTwoPi, the function forms two full circles, which correspond to the \PfZero and \PfZero[1500] isocalars. Thus, the coupling of \Ppi[1800] to both isoscalars can directly be deduced. In order to examine the result, we compare the corresponding phase of $\pi^{-}\pi^{+}$ (measured \wrt to \wave{1}{++}{0}{+}{\Prho}{S}) to an analog analysis using decays of $D_{s}\to K\pi\pi$ from Babar~\cite{Aubert:2008ao}, shown in \Cref{fig:deisobarred}c. To help the comparison, we use an arbitrary phase offset for our data. We can deduce that the phase of the \pipiSW shows a similar behaviour, whether originating from the three-body weak decay of $D_{s}$ mesons or from the strong decay of \PpiZero[1800], itself extracted by means of PWA from our data.

\section{Conclusion}
\label{sec:conclusion}
COMPASS has performed a two-step PWA on a data set comprising about $50\cdot 10^{6}$ events in the \threePi final state, using the so far largest model space with 88 waves. We have extracted resonance parameters for eleven isovector mesons with $C=+1$. 
%modeling the spectral function of partial waves by a coherent sum of relativistic Breit-Wigner amplitudes and a non-resonant contribution. 
The analysis revealed a new \PaOne[1420], the nature of which is much debated in the literature. We have confirmed the existence of three different $\pi_{2}$ states as well as \PpiOne. The width of \PpiOne is considerably larger than previous observations, although systematic uncertainties are large. Also \PaOne[1640] and \PaTwo[1700] have well been identified, though resonance parameters are still uncertain. This work constitutes the first coherent observation of all these states within one analysis. This allowed to study a large variety of systematic uncertainties for the resonance parameters connected to wave selection and model variation in these fits. 

\appendix
%\section{}

%Use the \verb|\appendix| command if you need an appendix(es). The \verb|\section| command should follow even though there is no title for the appendix (see above in the source of this file).
% ------------------------------------------------------------------------------
% bibliography
\bibliographystyle{unsrt}
\bibliography{../c3pipaper/trunk/spectroscopy}

%\begin{thebibliography}{9}
%\bibitem{jpsj} The abbreviation for JPSJ must be ``J. Phys. Soc. Jpn." in the reference list.
%\bibitem{instructions} More abbreviations of journal titles are listed in ``Instructions for Preparation of Manuscript", which is available at our Web site (http://jpsj.ipap.jp).
%\bibitem{format} F. Author, S. Author, and T. Author: Abbreviated journal title \textbf{volume in bold face} (year of publication) initial page or article ID.
%\end{thebibliography}

\end{document}